%% file: main.tex
\documentclass[journal]{IEEEtran}
\usepackage{enumitem}
\usepackage{amsmath, amssymb,bm}
\usepackage{booktabs}
\usepackage{graphicx,epstopdf,subfigure}
\usepackage{multirow}
\usepackage{tabularx,threeparttable}
\usepackage{cite}
\usepackage{array}
\usepackage{xcolor}
\usepackage{comment}

\usepackage[normalem]{ulem} 
\usepackage[colorlinks=true,citecolor=blue,urlcolor=blue]{hyperref}
\usepackage{mathtools}
\usepackage{graphicx,epstopdf,subfigure}
\usepackage{bbm}
\usepackage{threeparttable}
\usepackage{bm}
\usepackage[ruled]{algorithm2e}
\usepackage{amssymb}
\usepackage{pifont}

\newcommand{\cmark}{\text{\ding{51}}}%
\newcommand{\xmark}{\text{\ding{55}}}
\newcommand{\reviseminor}[1]{\textcolor{black}{#1}}
\newcommand{\revise}[1]{\textcolor{black}{#1}}
\ifCLASSINFOpdf
\else
\fi
\usepackage{fancyhdr}

\def\eg{\textit{e.g.}}
\def\ie{\textit{i.e.}}
\def\etal{\textit{et al.}}

\begin{document}


\title{MS-Net: Multi-Site Network for Improving Prostate Segmentation with Heterogeneous MRI Data}

\author{Quande Liu,~\IEEEmembership{Student Member,~IEEE}, Qi Dou,~\IEEEmembership{Member,~IEEE}, Lequan Yu,~\IEEEmembership{Student Member,~IEEE},\\ and Pheng Ann Heng,~\IEEEmembership{Senior Member,~IEEE}

\thanks{Manuscript received on 19 Aug 2019, revised on 21 Dec 2019,  accepted on 14 Feb 2020. This work was partially supported by a grant from Hong Kong Research Grants Council (Project No. CUHK 14225616) and a grant from the National Natural Science Foundation of China (Project No. U1813204).
\emph{(Corresponding author: Qi Dou.)}}
\thanks{Q. Liu, Q. Dou, L. Yu, and P. A. Heng are with the Department of Computer Science and Engineering, The Chinese University of Hong Kong, HK, China (emails: \{qdliu;~qdou\}@cse.cuhk.edu.hk, ylqzd2011@gmail.com, pheng@cse.cuhk.edu.hk).}
\thanks{L. Yu is also with Department of Radiation Oncology, Stanford University, Stanford, CA 94305 USA.}
\thanks{P. A. Heng is also with Shenzhen Key Laboratory of Virtual Reality and Human Interaction Technology, Shenzhen Institutes of Advanced Technology, Chinese Academy of Sciences, China.}
}
\markboth{IEEE Transactions on Medical Imaging}
{Shell \MakeLowercase{\textit{et al.}}: Bare Demo of IEEEtran.cls for Journals}

\maketitle

\thispagestyle{fancy}
\fancyhead{}
\lhead{}
\lfoot{}
\cfoot{\small{Copyright \copyright~2019 IEEE. Personal use of this material is permitted. However, permission to use this material for any other purposes must be obtained from the IEEE by sending a request to pubs-permissions@ieee.org.}}

\input{abstract.tex}

\input{introduction.tex}

\input{relatedwork.tex}
\input{method.tex}
\input{experiments.tex}

\input{discussion.tex}

\input{conclusion.tex}

\bibliographystyle{IEEEtran}
\small\bibliography{refs}

\end{document}

%% file: abstract.tex
\begin{abstract}
Automated prostate segmentation in MRI is highly demanded for computer-assisted diagnosis. 
Recently, a variety of deep learning methods have achieved remarkable progress in this task, usually relying on large amounts of training data.
Due to the nature of scarcity for medical images, it is important to effectively aggregate data from multiple sites for robust model training, to alleviate the insufficiency of single-site samples.
However, the prostate MRIs from different sites present heterogeneity due to the differences in scanners and imaging protocols, raising challenges
for effective ways of aggregating multi-site data for network training.
In this paper, we propose a novel multi-site network (MS-Net) for improving prostate segmentation by learning robust representations, leveraging multiple sources of data. 
To compensate for the inter-site heterogeneity of different MRI datasets, we develop Domain-Specific Batch Normalization layers in the network backbone, enabling the network to estimate statistics and perform feature normalization for each site separately. 
Considering the difficulty of capturing the shared knowledge from multiple datasets, a novel learning paradigm, \ie, Multi-site-guided Knowledge Transfer, is proposed to enhance the kernels to extract more generic representations from multi-site data. 
Extensive experiments on three heterogeneous prostate MRI datasets demonstrate that our MS-Net improves the performance across all datasets consistently, and outperforms state-of-the-art methods for multi-site learning. 

\end{abstract}


%% file: introduction.tex
\section{Introduction}

\IEEEPARstart{P}{rostate} diseases (e.g., prostate cancer, prostatitis and benign prostate hyperplasia) are common afflictions in men. Accurate segmentation of the prostate from the magnetic resonance image (MRI) is crucial for diagnosis and treatment planning of these diseases. 
Recently, convolutional neural networks (CNNs) have made remarkable progress for automated prostate segmentation~\cite{milletari2016vnet,yu2017volumetric,jia2019hdnet}. 
For example, Milletari~\etal~\cite{milletari2016vnet} proposed a V-shaped fully convolutional network (FCN) with Dice loss for accurate prostate segmentation.
However, a high performance of these data-driven CNN models commonly rely on large amounts of training samples. 
With the nature of data scarcity in medical imaging, such assumption is usually violated in real-world clinical practice, as it is hard to collect extensive samples from a single site (or hospital).
In such case, it is very meaningful to aggregate data that are acquired from multiple sites for robust model training.

\begin{figure}[t]
	\centering
	\revise{
	\includegraphics[width=0.47\textwidth]{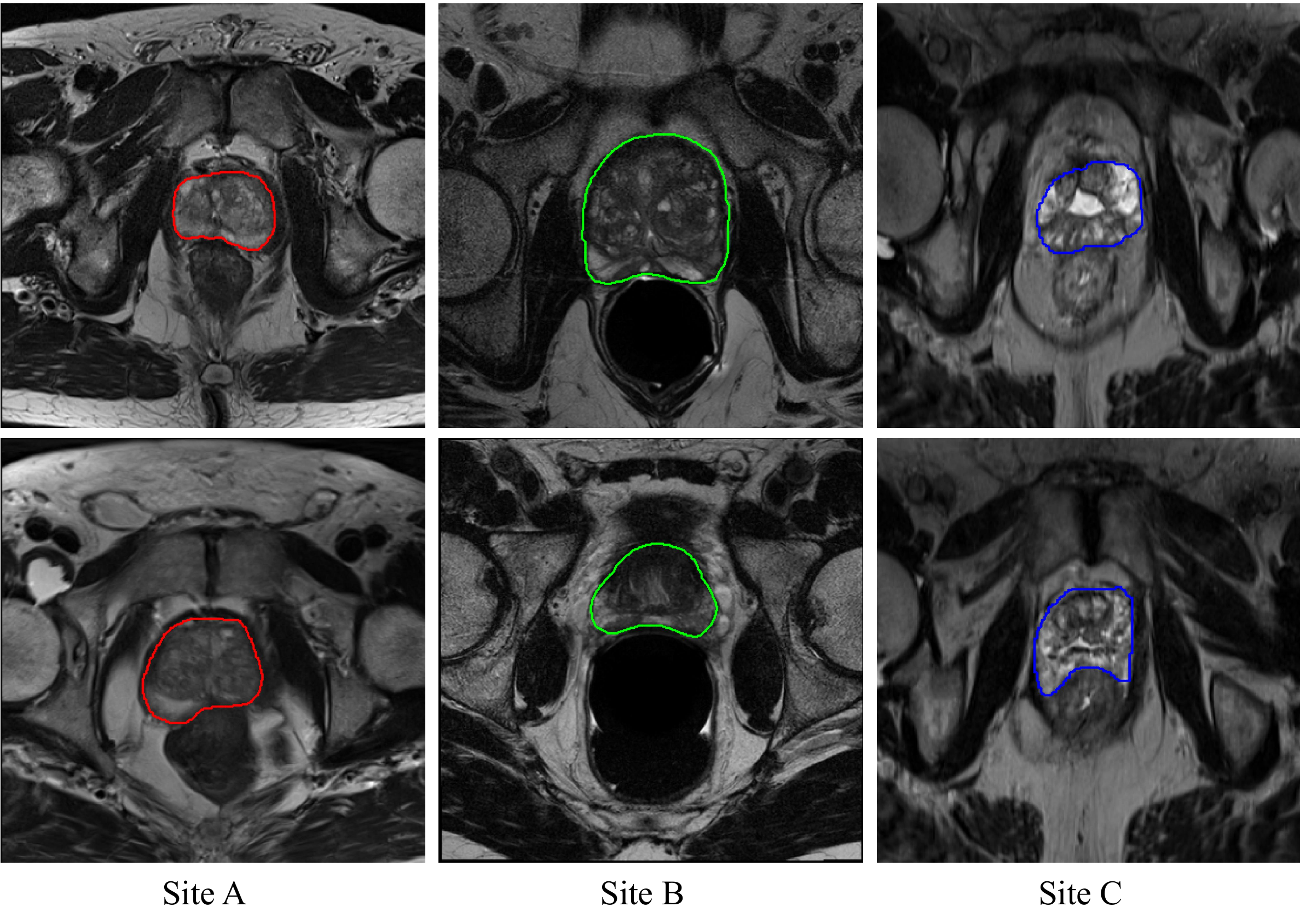}
	\vspace{-3mm}
	\caption{Normalized prostate MRI data from three different sites, showing the appearance differences. From left to right are samples from Radboud University Nijmegen Medical Centre (site A), Boston Medical Center (site B), and Hospital Center Regional University of Dijon-Bourgogne (site C), respectively.}
	}
	\label{fig:dataset}
	\vspace{-5mm}
\end{figure}

However, \revise{the prostate MRI data from different sites present apparent inter-site heterogeneity due to the differences in imaging protocols, endorectal coil usages, or population demographics, see Fig.~\ref{fig:dataset}.} These discrepancies inherent in multi-site data raise challenges when combining different sites of samples for model training, because a good performance of neural networks usually requires a well normalized data distribution~\cite{wang2019universalobject}. 
Previous works~\cite{eli2018intersite,styner2002multi-site, nie2013multisite} indicate that \reviseminor{directly mixing multi-site data for model training even though might improve the generalization performance on external sites, brings marginal improvements, or even leads to performance degradation on internal sites if the inter-site heterogeneity is significant}. 
John~\etal~\cite{john2019generali} aggregate two different prostate MRI datasets, and present that augmenting training data with heterogeneous images cannot bring explicit improvement. 
Rundo~\etal~\cite{rundo2019cross} conduct extensive experiments on \emph{inter-site} prostate MRIs, and find that directly mixing two datasets leads to performance degradation. 
Similar observations have also been reported on other medical image analysis tasks~\cite{john2018variable,kevin2019baseline}.
John~\etal~\cite{john2018variable} conduct cross-sectional study for pneumonia detection in chest radiographs, showing that a model trained with combined data is inferior to a model trained on individual site.

Therefore, towards this crucial real-world problem, effectively learning a robust and accurate segmentation model by taking advantage of multiple sources of data is non-trivial yet unsolved.
In this paper, we define this problem setting as \emph{multi-site learning}. Specifically, given a set of training datasets from $S$ different sites, we aim to jointly learn a single network from these $S$ datasets, and boost the segmentation performance on all the $S$ sites consistently, by leveraging the task knowledge shared across multiple sites. \reviseminor{Note that the performance on external sites is not considered in the scope.}
In multi-site learning, we define the \emph{Joint} approach as directly mixing multi-site data for network training and the \emph{Separate} approach as training a separate network on each site. 
The \emph{Joint} and \emph{Separate} models are the two strong baselines which we consider and compare, in addition to other multi-site learning methods. 

In this paper, we aim at conducting prostate segmentation from heterogeneous multi-site MRI data. 
We develop a novel multi-site network (MS-Net) for improved prostate segmentation by learning generic representations from multi-site data. 
To compensate for the inter-site heterogeneity, we propose a Domain-Specific Batch Normalization (DSBN) layer in the network backbone, enabling the network to estimate statistics and perform feature normalization for each site individually.
%
%
Based on previous observations~\cite{rundo2019cross}, it is not straight-forward for a network to explore robust representations from multiple datasets. Therefore, we propose a novel learning paradigm, \ie, Multi-site-guided Knowledge Transfer (MSKT) to enhance the learning of shared kernels by conducting knowledge transfer from multiple auxiliary branches.
%
%
Compared with conventional supervision only with ground truth, our MSKT provides richer multi-site information to regularize the shared kernels to capture more robust representations for more accurate segmentation. Extensive experiments demonstrate that compared with \emph{Joint/Separate} approaches, our MS-Net brings consistent and significant improvements on three heterogeneous MRI datasets.
Our main contributions are summarized as follows:
\begin{enumerate}
	\item
	We develop a novel framework of MS-Net for improving prostate segmentation with multi-site MRIs. Our model explicitly compensates for the inter-site data heterogeneity with domain-specific batch normalization layers. 
	
    
    \item
    We propose a novel learning paradigm,~\ie~multi-site-guided knowledge transfer, to guide the shared CNN kernels to capture more robust representations from multi-site data.
	
	\item
	Extensive experiments using three prostate MRI datasets demonstrate that our approach consistently improves the segmentation accuracy on all three datasets, outperforming the strong baselines and the state-of-the-art multi-site learning approaches. \revise{Code of our method is publicly available at: \url{https://github.com/liuquande/MS-Net}.}
\end{enumerate}

In the remainders of this paper, we present the related works in Section~\ref{sec:related}, elaborate our proposed method in Section~\ref{sec:method}, and describe our extensive experimental results in Section~\ref{sec:experiment}. We then discuss and analyze our work in Section~\ref{sec:discussion} and draw a conclusion in Section~\ref{sec:conclusion}.

%% file: relatedwork.tex
\section{Related Works}
\label{sec:related}

\subsection{Multi-site Learning}
To improve the effectiveness of utilizing multiple sites of data, in medical image analysis, a variety of image intensity normalization methods~\cite{wei2004normalization, shi2014statistical} have been proposed to alleviate the inter-site appearance difference in the pre-processing step, and therefore to improve the performance on cross-site datasets.
Meanwhile, some previous works studied how to design effective hand-crafted features and classifiers for medical image analysis tasks across different domains~\cite{prath2018empirical, opbroek2018weighting,wang2015classification,ma2018classification,opbroek2015transfer}.
For example, Wang~\etal~\cite{wang2015classification} incorporated multiple classification tasks of bipolar disorder and extended the original support vector machine (SVM) for joint learning. Ma~\etal~\cite{ma2018classification} established a multi-task learning framework, constrained by sparsity regulating terms, to learn domain-shared and domain-specific features utilizing data collected from three different academic centers. 
Moreover, Opbroek~\etal~\cite{opbroek2015transfer} presented four transfer classifiers to employ datasets of different characteristics and 
outperformed common  supervised-learning approaches.
These works, however, relied heavily on the quality of hand-crafted features. 
Recently, some works utilized deep learning techniques to brige the inter-site variability~\cite{rundo2019use,karani2018lifelong,dou2020unpaired,chang2019domain}.
Rundo~\etal~\cite{rundo2019use} conducted channel-wise feature calibration to improve the prostate segmentation across heterogeneous datasests. 
To adapt a brain segmentation network to a new modality, Karani~\etal~\cite{karani2018lifelong} observed that fine-tuning the BN layers on the new modality while fixing other CNN weights was helpful to preserve the cross-site shared information.
\reviseminor{Similarly, Chang~\etal~\cite{chang2019domain} observed that independently normalizing features from different domains produced competitive performance under the setting of domain adaptation, which could demonstrate the effectiveness of independent feature normalization for handling domain shift problem.}

In nature image processing community, the multi-site learning is also related to multi-domain learning~\cite{rebuff2018efficient,wang2019universalobject}, which aims to develop a single network that could perform well on diverse visual domains, achieving higher parameters efficiency across tasks. 
These works highlighted that assigning domain-specific parameters, \eg, CNN kernels, was helpful to tackle domain-specific nuances and boost cross-domain generalization. 
However, multi-domain learning focus on studying how to develop a common network for handling different objects of largely distinct patterns (\eg, Aircraft vs. Flower) to avoid training multiple networks, while our work aims to leverage multiple similar datasets to improve the performance on different sites simultaneously, leaving more space to dig the potential of utilizing the shared information across different datasets. 

\begin{figure*}[t]
	\centering
	\includegraphics[width=\textwidth]{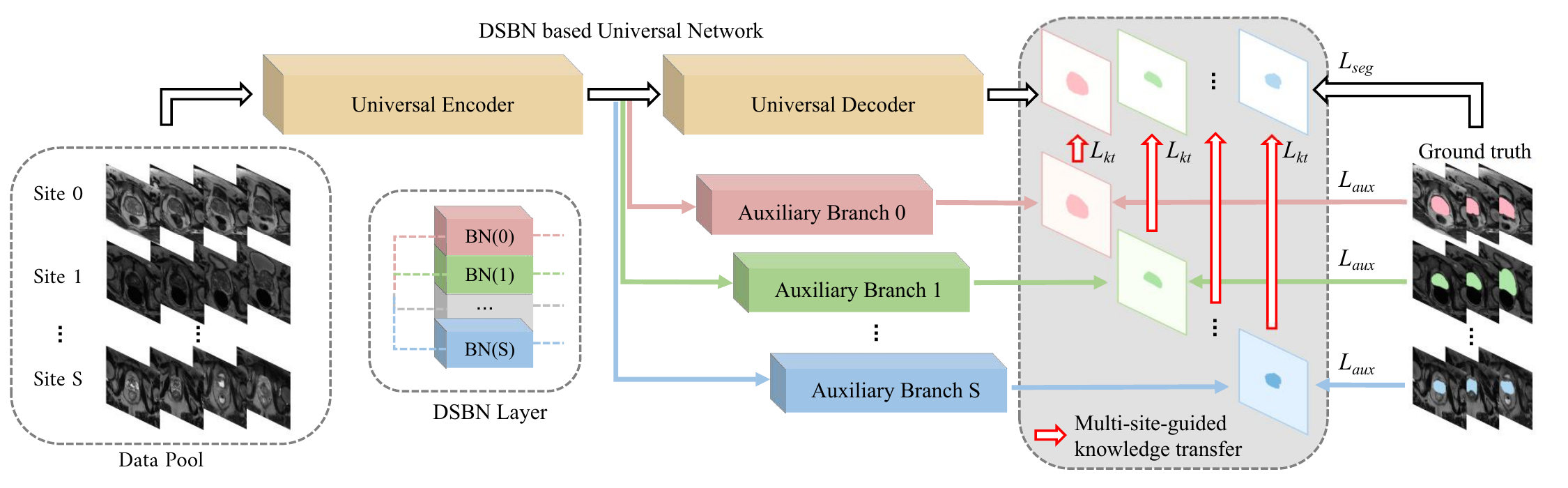}
	\vspace{-3mm}
	\caption{The schematic illustration of our proposed multi-site network (MS-Net), consisting of a universal network and $S$ domain-specific auxiliary branches. Both the encoder and decoder parts of universal network utilize the DSBN layer to compensate for the inter-site heterogeneity. In each iteration, we feed the framework with $S$ batches of images, one batch from each site. The universal network is trained synergistically with the supervision of ground truth and transferred multi-site knowledge from auxiliary branches to help explore the general representation shared across different datasets.} 
	\label{fig:overview}
	\vspace{2mm}
\end{figure*}

\subsection{Prostate Segmentation}
Previous automated prostate segmentation methods can be broadly categorized into three branches: multi-atlas-based methods\cite{klein2008automatic}, deformable methods\cite{toth2012multi} and learning-based methods\cite{milletari2016vnet}. 
For example, Klein~\etal~\cite{klein2008automatic} utilized atlas matching for prostate segmentation. The general idea is to register segmented template images with target images and then fused the aligned segmentations to derive the final results. 
Regarding deformable methods, Toth~\etal~\cite{toth2012multi} proposed an improved active appearance models (AAMs) to accurately model the shape information for prostate delineation. 
In addition, there are also some studies utilizing graph cut~\cite{tian2017super} and feature-based machine learning methods~\cite{zheng2014marginal} for prostate segmentation. 
Recently, more attention has been drawn to deep-learning-based methods~\cite{tian2018psnet,wang2019deeply,cheng2017automatic,clark2017fully,zhu2017deeply,wang2018automatic,zhang2019znet,wang2019deep,jia2020apa} due to its automatic representation learning capability. 
Milletari~\etal~\cite{milletari2016vnet} proposed a V-shape fully convolutional networks (FCN) for fast and accurate prostate segmentation, while Yu~\etal~\cite{yu2017volumetric} incorporated the residual connections and deep supervision mechanism into the 3D networks for improving the prostate segmentation. 
\revise{To improve the delineation of blurry boundary, Nie~\etal~\cite{nie2019semantic} developed a semantic-guided feature learning strategy to learn more discriminative representations. Zhu~\etal~\cite{zhu2019boundary} developed a boundary-aware adversarial learning strategy to improve the boundary delineation on source domain by introducing external datasets.
However, all the above works either directly utilize single-site or multi-site prostate MRIs for network training, without carefully analyzing the inter-site heterogeneity, or differentiate data from different sites but only focus on improving the performance on a certain data site. Differently, this work focuses on a more general setting,~\ie, multi-site learning, which aims to tackle the data heterogeneity of multi-site data and effectively leverage multi-site data to learn more robust representation for improved prostate segmentation on multiple sites simultaneously. }

%% file: method.tex
\section{Methodology}
\label{sec:method}
The Fig.~\ref{fig:overview} is an overview of our proposed multi-site network (MS-Net) for prostate segmentation. Our MS-Net consists of a \emph{universal network} and several \emph{auxiliary branches} which are associated with specific domains.
As setting of multi-site learning, we have a set of data collected from $S$ different sites. 
%
The universal network utilizes Domain Specific Batch Normalization (DSBN) layers to tackle the inter-site heterogeneity. The synergistical guidance of ground truth label and the transferred multi-site knowledge from auxiliary branches facilitates the shared CNN kernels in universal network to learn more robust representation from multi-site data.
\subsection{Domain Specific Batch Normalization Layer}
Batch normalization~\cite{ioffe2015batch} has been widely used in CNNs for reducing internal covariate shift, and therefore helping improve feature discrimination capability and speed up learning process.
Its central idea is to normalize the internal representations along the channel dimension, then apply affine transformation on the whitened feature maps with trainable parameters $[\gamma, \beta]$. 
Let $x_k\in[x_1, \dots , x_K]$ be a certain channel of the K-channel feature maps in a certain layer, the corresponding normalized representations $y_k \in [y_1, \dots, y_K]$ are computed as:
\begin{equation}
y_k = \gamma \cdot \hat{x_k} + \beta,\quad \text{where} \quad \hat{x_k}=\frac{x_k- E[x_k]}{\sqrt{Var[x_k]}+\epsilon},
\end{equation}
where $E[x]$ and $Var[x]$ are the mean and variance of $x$, and $\epsilon$ is an infinitesimal. 
Meanwhile, the BN layer collects a moving mean and a moving variance in the training process to capture the global statistics, and utilizes these estimated moving values to conduct feature normalization in the testing phase. 

\begin{figure}[t]
	\centering
	\includegraphics[width=0.48\textwidth]{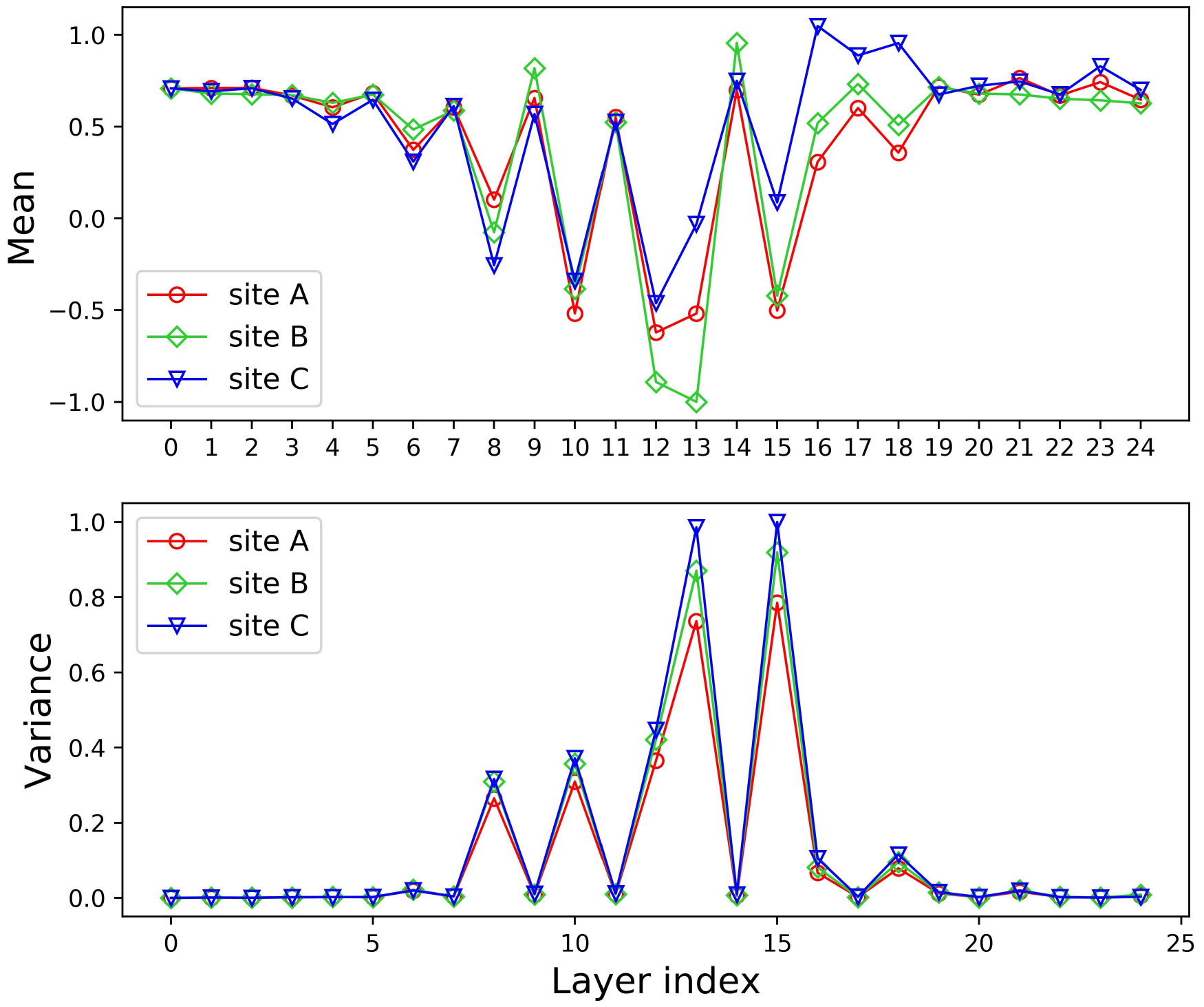}
	\vspace{-1mm}
	\caption{The global statistics (BN moving mean and moving variance) collected by single-site segmentation model trained on each site. The mean and variance values are the average of BN statistic for each layer. Note that all models are trained from the same initialization state and training hyper-parameters.}
	\label{fig:bn}
	\vspace{-2mm}
\end{figure}

In our problem setting, the prostate MRI images are acquired from different sites using various scanners and imaging protocols. 
In Fig.~\ref{fig:bn},
we visualize the single-site global statistics (collected moving mean and moving variance) captured by BN layers of the separate networks trained on each site individually.
It is observed that these statistics present visible variations across different sites, especially for those middle layers where the features are high-dimensional with more channels (cf. Table~\ref{tab:network}).
Based on these observations, we argue that directly integrating all heterogeneous datasets for training is not effective due to two reasons: 
1) The statistical difference of heterogeneous datasets might 
bring difficulty for learning generic representations, since the shared kernels would bother with the nonessential domain-specific variations; 
2) The BN layers may result in inaccurate estimation of global mean and variance in the training phase given multi-site statistical differences. Directly sharing the estimated values in the testing phase would lead to performance degradation.

To overcome these limitations, we incorporate a Domain-Specific Batch Normalization (DSBN) layer into the universal network, by assigning an individual BN layer for each site to effectively tackle the inter-site discrepancy. %
Specifically, the DSBN layer allocates each site $s$ with domain-specific trainable variables $[\gamma^s, \beta^s]$.
Let $x_k^s\in[x_1^s, \dots, x_K^s]$ be a certain channel of feature maps of a sample from site $s$, the corresponding output $y_k^s$ is expressed as:
\begin{equation}
y_k^s = \gamma^s \cdot \hat{x}^s_{k}  + \beta^s,\quad \text{where} \quad \hat{x}_k^s =\frac{x_k^s-E[x_k^s]}{\sqrt{Var[x_k^s]+\epsilon}}.
\end{equation}
In testing phase, the DSBN layer applies the collected accurate domain-specific moving mean and moving variance for future normalization to data from the corresponding site.
%
Compared with sharing BN layers across different sites, the DSBN supplies domain-specific variables to handle domain-specific nuances and task-irrelevant inter-site variations by performing individual feature normalization.
%

\subsection{Multi-site-guided Knowledge Transfer}
When overcoming the inter-site variances, our goal is to extract more robust representations by aggregating multi-site data. There are many ways for a network to converge to zero training error, while some way generalizes better than the others due to being in ``wide valleys", ranther than in sharp minima~\cite{chaud2017entropy}. Under multi-site learning, it is obsered that the universal network rapidly converges to the sharp minima of each dataset under conventional supervised learning, instead of exploring a robust global minima among multiple datasets~\cite{rundo2019cross}. 

With these insights, we propose a novel learning paradigm, \ie, Multi-site-guided Knowledge Transfer (MSKT), to enhance the domain-agnostic CNN kernels to dig the ``wide valley" among different datasets and therefore capture more robust representations. Recent study on lifelong learning~\cite{hou2018lifelong} shows that compared with learning directly from ground truth label, learning from intermediate CNNs through knowledge distillation provides a shortcut for a network to explore the generic representations between different tasks.

As shown in Fig.~\ref{fig:overview}, we synergistically train our universal network with the supervision from ground truth labels and additional multi-site knowledge from auxiliary branches.
Specifically, we incorporate $S$ domain-specific auxiliary branches into the universal network, each of which has the same architecture as the universal decoder.
The auxiliary branches serve as an independent feature extractor for each site and could learn the specific knowledge from each site more comprehensively compared with the universal network.
Each auxiliary branch $A^{s}$ is trained with conventional segmentation dice loss~\cite{milletari2016vnet}. 

Concurrently, we transfer the multi-site knowledge from auxiliary branches into the universal network with an effective knowledge transfer scheme. 
In each iteration, instead of transferring the knowledge from each auxiliary branch sequentially into universal network, we conduct knowledge transfer from all auxiliary branches together into universal network, encouraging the shared kernels in universal network to capture more generic representations. 
Specifically, the total objective function to train the universal network with site $s$ consists of a conventional supervised dice loss $L^s_{\text{uni}}$ and a \textit{knowledge transfer loss} $L^s_{\text{kt}}$. 
Different from the previous knowledge distillation methods~\cite{hinton2014distill,zhang2018deep}, \revise{we adopt a dice-coefficient-like knowledge transfer loss to align the probability maps of universal network with the segmentation masks from auxiliary branches. Note that we transform the segmentation masks into one-hot format to keep the dimensions consistent with the probability maps.
Denote the one-hot prediction masks of auxiliary branches as $P_{\text{aux}}^s \in \mathbb{R}^{b \times h \times w \times c}$, the activation values following the softmax layer of universal network as $M_{\text{uni}}^s \in \mathbb{R}^{b \times h \times w \times c}$, where $b$ is the batch size, $h$ and $w$ are the spatial dimensions of feature map, $c$ is the channel number.}
The knowledge transfer loss can be computed as follows:
\begin{align}
L_{\text{kt}}^s(M_{\text{uni}}^s,P_{\text{aux}}^s) &= 1-\frac{2 \sum_i^\Omega m_i^s\cdot p_i^s}{\sum_i^\Omega (m_i^s)^2 + \sum_i^\Omega (p_i^s)^2},
\end{align}
where $m_i^s$$\in$$M_{\text{uni}}^s$, $p_i^s$$\in$$P_{\text{aux}}^s$, $\Omega$ denotes the total number of pixels in one batch. 

Intuitively, our idea of MSKT is in line with recent studies on the robustness of high posterior entropy~\cite{chaud2017entropy, pereyra2017regulariz}.
In our MSKT, each auxiliary branch only aims to better learn the specific knowledge from a certain dataset. They learn different representations and produce different predictions compared with the universal network. These factors provide extra multi-site information for the training of universal network. 
Under conventional supervised learning, the universal network converges quickly if the network capacity is large. 
While in MSKT, the universal network has to mimic the ground truth label and the predictions of multiple auxiliary branches simultaneously.
Compared with conventional supervised learning, MSKT provides additional multi-site information to regularize the universal network and increases its posterior entropy~\cite{chaud2017entropy}, which helps the shared kernels to explore 
more robust representation among multiple datasets. 
Moreover, the multi-branch architecture in MSKT could also perform as a positive feature regularization to the universal encoder by jointly training the auxiliary branches and the universal network. 
\begin{table}[t]
\centering
\caption{Architecture of the network backbone.}
\label{tab:network}
\begin{tabular}{c|c|c}
\hline
                 & Feature size & Layer                                                                                       \\ \hline
input            & 384x384      &                                                                                             \\ \hline
convolution 1    & 384x384      & 3x3, 32, stride 2                                                                           \\ \hline
residual block 1 & 384x384      & \begin{tabular}[c]{@{}c@{}}3x3, 32  conv\\ 3x3, 32  conv\end{tabular}                       \\ \hline
pooling 1        & 192x192      & 3x3 max pool, stride 2                                                                      \\ \hline
residual block 2 & 192x192      & \begin{tabular}[c]{@{}c@{}}3x3, 64  conv\\ 3x3, 64  conv\end{tabular}                       \\ \hline
pooling 2        & 96x96        & 3x3 max pool, stride 2                                                                      \\ \hline
residual block 3 & 96x96        & \begin{tabular}[c]{@{}c@{}}3x3, 128 conv\\ 3x3, 128  conv\end{tabular}                      \\ \hline
pooling 3        & 48x48        & 3x3 max pool, stride 2                                                                      \\ \hline
residual block 4 & 48x48        & \begin{tabular}[c]{@{}c@{}}3x3, 256  conv\\ 3x3, 256  conv\end{tabular}                     \\ \hline
pooling 4        & 24x24        & 3x3 max pool, stride 2                                                                      \\ \hline
residual block 5\_1 & 24x24        & \begin{tabular}[c]{@{}c@{}}3x3, 512  conv\\ 3x3, 512  conv\end{tabular}                     \\
\hline
residual block 5\_2 & 24x24        & \begin{tabular}[c]{@{}c@{}}3x3, 512  conv\\ 3x3, 512  conv\end{tabular}                     \\
\hline \hline
upsample 6       & 48x48        & \begin{tabular}[c]{@{}c@{}}3x3, 256 deconv- {[}res block 4{]}\\ 3x3, 256, stride 2\end{tabular} \\ \hline
residual block 6 & 48x48        & \begin{tabular}[c]{@{}c@{}}3x3, 256  conv\\ 3x3, 256  conv\end{tabular}                     \\ \hline
upsample 7       & 96x96        & \begin{tabular}[c]{@{}c@{}}3x3, 128 deconv- {[}res block 3{]}\\ 3x3, 128, stride 2\end{tabular} \\ \hline
residual block 7 & 96x96        & \begin{tabular}[c]{@{}c@{}}3x3, 128  conv\\ 3x3, 128  conv\end{tabular}                     \\ \hline
upsample 8       & 192x192      & \begin{tabular}[c]{@{}c@{}}3x3, 64 deconv- {[}res block 2{]}\\ 3x3, 64, stride 2\end{tabular}   \\ \hline
residual block 8 & 192x192      & \begin{tabular}[c]{@{}c@{}}3x3, 64  conv\\ 3x3, 64  conv\end{tabular}                       \\ \hline
upsample 9       & 384x384      & \begin{tabular}[c]{@{}c@{}}3x3, 32 deconv- {[}res block 1{]}\\ 3x3, 32, stride 2\end{tabular}   \\ \hline
residual block 9 & 384x384      & \begin{tabular}[c]{@{}c@{}}3x3, 32  conv\\ 3x3, 32  conv\end{tabular}                       \\ \hline
output 10    & 384x384      & 1x1, 2 conv                                                                                     \\ \hline
\end{tabular}
\end{table}
\subsection{Network Architecture and Implementation Details}
\subsubsection{Network backbone}
The choice of network backbone in our proposed multi-site learning method is flexible.
Most recent networks on medical image segmentation can be employed. We adopt an adapted 2D Residual-UNet~\cite{yu2017volumetric} as the segmentation network backbone, which achieves remarkable performance in the prostate segmentation problem. 
Due to the large variance on the through-plane resolution among data from different sites, we employ the 2D network architecture. 
The detailed structure of the adapted network backbnone is shown in Table~\ref{tab:network}, which contains 4 down-sample and 4 up-sample blocks. 
Note that ``3x3, 32 conv/deconv" denotes the sequence DSBN-Relu-Conv/Deconv layers with kernel size 3x3 and output channel 32. The symbol ``-[*]" denotes the long range summation connection with the output of ``*" block. 
In the universal network, all normalization layers was replaced to DSBN layers to handle the inter-site discrepancy. 
Each auxiliary branch has the same architecture as the decoder part of the universal network, but substitutes the DSBN layers with BN layers. 
\begin{algorithm}[t]
\caption{Training procedure of the proposed MS-Net}\label{algorithm}
\KwData{Datasets $D_1$, \dots, $D_S$ from from $S$ different sites, Training iteration $\tau$}
\KwResult{Universal encoder $\bm{\theta}_e$, Universal decoder $\bm{\theta}_d$, Auxiliary branch $\{\bm{\theta}_i\}_{i=1}^S$;}
|\textbf{Training}|\\
\textbf{Initialization:} t=1; Randomly initialize $\bm{\theta}_e$,$\bm{\theta}_d$,$\{\bm{\theta}_i\}_{i=1}^S$\;
\While{$t \leq \tau$}
{Given $S$ mini-batches from $\{D_i\}_{i=1}^S$ \;
Compute the loss function of all auxiliary branches $L_{\text{aux}}$ in Eq.~(\ref{eq:totalloss1})\; 
Update parameters $\bm{\theta}_e$ and $\{\bm{\theta}_i\}_{i=1}^S$\;
\BlankLine
Compute the loss function of universal network $L_{\text{uni}}$ in Eq.~(\ref{eq:totalloss1})\;
Update parameters $\bm{\theta}_e$ and $\bm{\theta}_d$\;}
|\textbf{Testing}|\\
Leave $\{\bm{\theta}_i\}_{i=1}^S$ and only keep $\bm{\theta}_e$, $\bm{\theta}_d$ for deployment.
\end{algorithm}
\subsubsection{Objective functions and training procedure}
The overall objective functions for updating auxiliary branches and the universal network are:
\begin{align}
\label{eq:totalloss1}
L_{\text{aux}} &= \sum^S_{s=1} L_{\text{aux}}^s + \eta(||\theta _e||_2^2 + \sum^S_{s=1} ||\theta _{\text{aux}}^s||_2^2), \\ \nonumber
\label{eq:totalloss2}
L_{\text{uni}} &= \sum^S_{s=1}{(\alpha L_{\text{kt}}^s + (1-\alpha )L_{\text{uni}}^s)} + \eta(||\theta _e||_2^2 + ||\theta _d||_2^2),
\end{align}
where $L_{\text{aux}}^s$ and $L_{\text{uni}}^s$ are segmentation dice loss for auxiliary branches and universal network, $L_{\text{kt}}^s$ is knowledge transfer loss for universal network. 
The $\alpha$ is a hyper-parameter to balance the segmentation loss and the knowledge transfer loss, which was set as 0.5 in our implementation and we also study this parameter in ablation study. 
The other terms are  $L2$ regularization, where the $\{\theta_e, \theta_d, \theta_{\text{aux}}^s\}$ are trainable parameters of universal encoder, universal decoder and auxiliary branches, respectively. %
The weight $\eta$ was set as $1e^{-4}$.

The training procedure is summarized in Algorithm~\ref{algorithm}. 
In our approach, the knowledge transfer is performed throughout the whole training process. 
For each training iteration, we feed the network with $S$~batches of images, each batch from one dataset.
The auxiliary branches and the universal network are trained alternatively. 
Once the training is finished, we remove all auxiliary branches and only preserve the universal network for inference. 
%

\subsubsection{Implementation details}
Our framework was implemented in Python with TensorFlow using three NVIDIA TitanXp GPUs. The computation for each site is conducted on one GPU. 
The network was trained using Adam optimizer with $\beta_1$=0.9, $\beta_2$=0.999. 
We totally trained 30000 iterations and the batch size was set as 5.
The learning rate was initialized as $1.0\times10^{-3}$ and decayed with a power of 0.95 after ever 500 iterations. \revise{Data augmentation of random horizontal flipping and random shift are used to mitigate the overfitting problem.} 
In post-processing, we conducted the morphological operation \revise{in 3D to select the largest connective volume} as the final segmentation mask.
    

%% file: experiments.tex
\section{Experiments}
\label{sec:experiment}
\subsection{Datasets and Evaluation Metric}
We collect prostate T2-weighted MRI from three different sites to evaluate the performance under multi-site learning, including 30 samples from Radboud University Nijmegen Medical Centre (Site A), 30 samples from Boston Medical Center (Site B) and  19 samples from Hospital Center Regional University of Dijon-Bourgogne (Site C).
Among these data, samples of Site A and Site B are from \emph{NCI-ISBI 2013 challenge (ISBI 13)} dataset~\cite{bloch2019isbi}, samples of Site C are from \emph{Initiative for Collaborative Computer Vision Benchmarking (I2CVB)} dataset~\cite{lemaitre2015i2cvb}. 

\begin{figure}[t]
	\centering
	\includegraphics[width=0.48\textwidth]{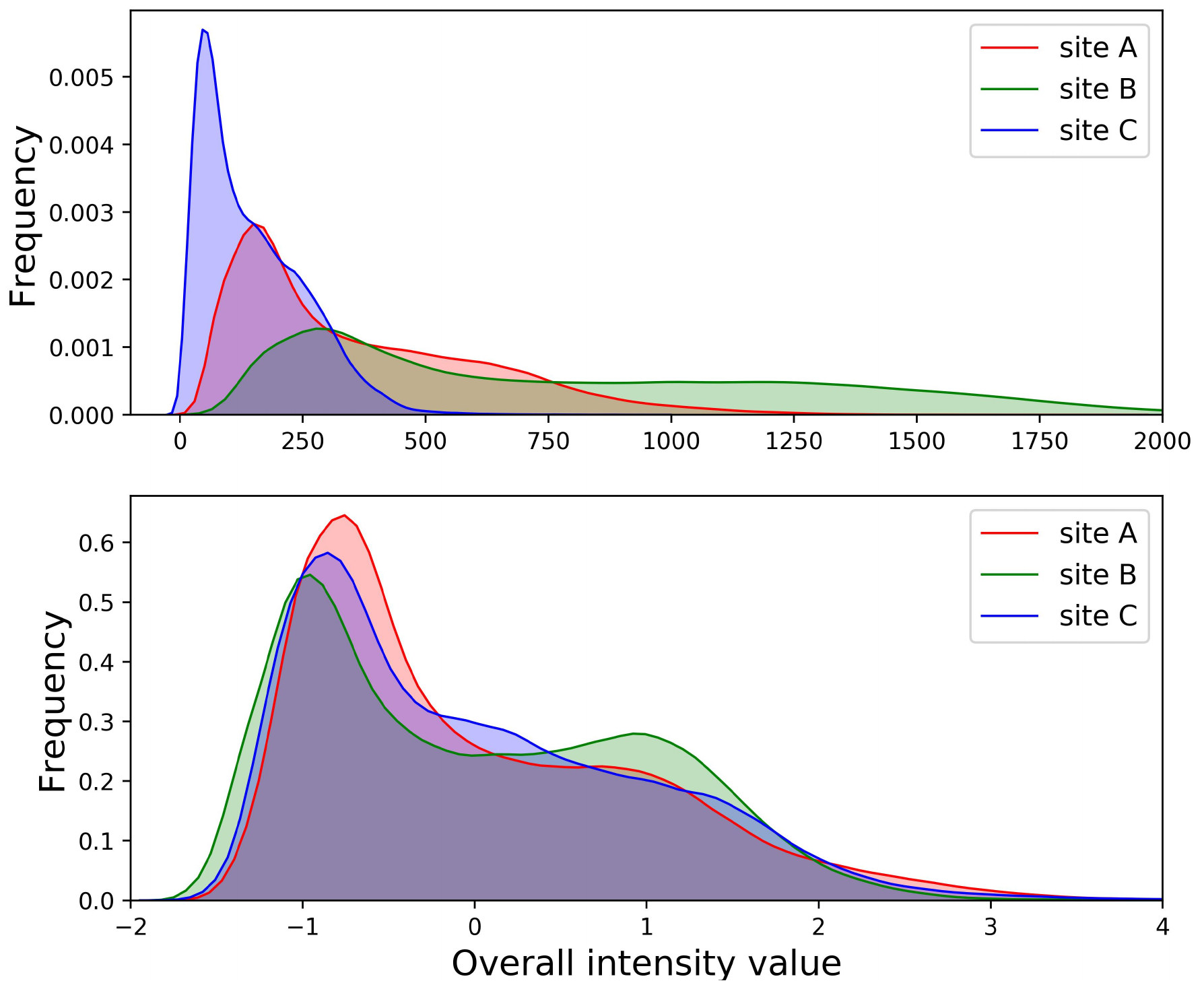}
	\vspace{-1mm}
	\caption{The overall intensity histogram distributions of the data from each site before (top) and after (bottom) intensity normalization. We normalize the intensity of each patient volume to zero mean and unit variance.}
	\label{fig:statistic}
	\vspace{-2mm}
\end{figure}

Data from three different sites are sampled using different acquisition protocols, as summarized in Table~\ref{table:dataset}.
The differences in scanners, field strength and coil type lead to large inter-site discrepancy. Meanwhile, images from different sites present heterogeneous in-plane and through-plane resolution, see the fourth column in Table~\ref{table:dataset}. 
In addition, different from the other two datasets, most images (17 out of 19) in Site C are acquired from patients with prostate cancer, leading to a semantic difference in the prostate area. These differences above lead to a visible appearance difference among multi-site data, see Fig.~\ref{fig:dataset}; and also lead to an intensity distribution shift, see Fig.~\ref{fig:statistic} (a).
\begin{table}[t]
    \renewcommand\arraystretch{1.2}
    \centering
    \caption{Details of the scanning protocols for three different sites.}
    \label{table:dataset}
    \begin{tabular}{lm{0.6cm}m{0.8cm}m{2cm}m{1cm}m{1cm}}
    \toprule
    Dataset & Case num      & Field strength (T) & Resolution(in-plane/through-plane)(mm) & Coil & Manufactor \\
    \hline
    Site A & 30  & 3 & 0.6-0.625/3.6-4 & Surface & Siemens  \\
    Site B & 30  & 1.5& 0.4/3 & Endorectal& Philips   \\
    Site C & 19  & 3 & 0.67-0.79/1.25   & No & Siemens \\
    \hline
    \end{tabular}
\end{table}
\begin{table}[t]
    \renewcommand\arraystretch{1.2}
	\centering
    \revise{
	\caption{\small{Comparison of joint approach with different pre-processing techniques.}}
	\label{tab:pre-processing}
	\scalebox{0.90}{
	\begin{tabular}{l| ccc | c  c c | c}
		\toprule
        Methods & BFC& NF& Intensities& Site A & Site B &Site C & Overall \\
        \hline
        Separate (A)&$\xmark$ &$\xmark$&whitening&90.47 &76.44 &56.81\\
        Separate (B)&$\xmark$ &$\xmark$&whitening&70.11 &90.52 &50.25 &90.56\\
        Separate (C)&$\xmark$ &$\xmark$&whitening&57.93 &55.25 &90.70\\
        \hline
        Joint  &$\xmark$ &$\xmark$&$\xmark$ &86.51 &88.00 &86.78 &87.10\\
        Joint  &$\xmark$ &$\xmark$&histogram &87.68 &88.02 &89.46 &88.39\\
        Joint  &$\xmark$ &$\xmark$&scaled &90.43 &88.06 &88.26 &88.92\\
        Joint  &$\xmark$ &$\xmark$&whitening &90.69 &89.53 &90.55 &90.25\\
        Joint  &$\xmark$ &$\cmark$&whitening &90.76 &89.46 &90.91 &90.37\\
        Joint  &$\cmark$ &$\xmark$&whitening &90.84 &89.81 &90.81 &90.49\\
        Joint  &$\cmark$ &$\cmark$&whitening &91.14 &89.75 &90.83 &90.58\\
		\bottomrule
	\end{tabular}
	}
	}
\end{table}

We conduct pre-processing for the three datasets. 
Following~\cite{rundo2019cross}, we first center-cropped the images from Site C with roughly same view as images from the other two sites, since the raw images of Site C are scanned from whole body, rather than prostate surrounding area. We then resized all samples of site A, B and C with size of $384 \! \times \! 384$ \revise{in axial plane. }
To reduce the intensity variance among different site samples, we normalized each sample to have zero mean and unit variance in intensity value before inputting to the network. 
%
\revise{We separate data from each site into 80\% and 20\% for training and testing. }
\revise{Evaluation metrics of Dice coefficient (\%) and Average symmetric distance (mm) are used to measure the segmentation performance in terms of whole object and surface.} We present the metric in the format of \emph{mean$\pm$std} to show the average performance as well as the cross-subject variance when comparing with other methods. 
%
\begin{table*}[h]
    \renewcommand\arraystretch{1.2}
	\centering
    \revise{
    	\caption{\small{Comparison on the segmentation performance of MS-Net and other state-of-the-art approaches.}}
    	\label{tab:comparisons}
    	\scalebox{1.00}{
    	\begin{tabular}{l|  c  c  c | c|  c  c  c | c}
    		\hline
            & \multicolumn{4}{c|}{\textbf{Dice Coefficient (mean$\pm$std, $\%$)}} & \multicolumn{4}{c}{\textbf{Average Symmetric Distance (mean$\pm$std, $mm$)}} \\ 
            \hline
            Methods&Site A &Site B &Site C & Overall &Site A &Site B &Site C & Overall\\
            \hline
            Tian~\etal~\cite{tian2017super} &88.23&88.23&---&\\
            Rundo~\etal~\cite{rundo2019cross} &---&---&88.66&\\
    		\hline
            Separate &90.47$\pm$3.00 &90.52$\pm$2.45 &90.70$\pm$3.34 &90.56$\pm$2.88 &1.02$\pm$0.42 &0.84$\pm$0.32 &0.75$\pm$0.33 &0.87$\pm$0.38\\
            Joint &90.69$\pm$3.05 &89.53$\pm$2.97&90.55$\pm$3.18 &90.25$\pm$3.08&0.96$\pm$0.37 &0.90$\pm$0.34 &0.75$\pm$0.32 &0.87$\pm$0.36\\
            \hline 
            USE-Net~\cite{rundo2019use} &90.90$\pm$2.41 &90.17$\pm$2.61 &90.73$\pm$2.36 &90.60$\pm$2.50 &0.90$\pm$0.32 &0.85$\pm$0.31 &0.70$\pm$0.30 &0.82$\pm$0.32 \\
            Dual-Stream~\cite{valindria2018multimodal} &90.87$\pm$2.85 &90.57$\pm$2.12 &90.10$\pm$3.28 &90.51$\pm$2.72 &0.92$\pm$0.38 &0.84$\pm$0.27 &0.75$\pm$0.32 &0.83$\pm$0.33\\
            Series-Adapter~\cite{rebuff2017learning} &90.80$\pm$2.72 &89.92$\pm$2.80 &91.24$\pm$2.21 &90.65$\pm$2.71 &0.95$\pm$0.42 &0.92$\pm$0.38 &0.71$\pm$0.28 &0.86$\pm$0.39\\
            Parallel-Adapter~\cite{rebuff2018efficient}  &90.61$\pm$3.54 &90.71$\pm$2.17 &91.30$\pm$2.06 &90.88$\pm$2.79 &0.96$\pm$0.25 &0.83$\pm$0.29 &0.74$\pm$0.28 &0.84$\pm$0.28\\
            \hline
            DSBN (ours) &90.98$\pm$2.69 &90.67$\pm$2.22 &91.07$\pm$1.86 &90.91$\pm$2.36 &0.95$\pm$0.48 &0.83$\pm$0.30 &0.74$\pm$0.24 &0.84$\pm$0.38\\
            \textbf{MS-Net (ours)}  &\textbf{91.54$\pm$2.01} &\textbf{91.24$\pm$1.97} &\textbf{92.18$\pm$1.62} &\textbf{91.66$\pm$1.95}  &\textbf{0.89$\pm$0.33} &\textbf{0.76$\pm$0.25} &\textbf{0.67$\pm$0.23} &\textbf{0.77$\pm$0.29}\\
            \hline
    	\end{tabular}
	}
	}
\end{table*}

\begin{table}[t]
    \renewcommand\arraystretch{1.2}
	\centering
	\caption{Paired t-Test for our method with two baseline methods using Dice Score.}
	\label{tab:test}
	\begin{tabular}{l|  c  c c c}
		\hline
        Methods&Site A & Site B &Site C &Overall\\
		\hline
        Joint &0.0115 &0.0002 &0.0036 &1.3e-7\\
        Separate &0.0083 &0.0013 &0.0105 &9.7e-6\\
		\hline
	\end{tabular}	
\end{table}
\begin{table}[t]
    \renewcommand\arraystretch{1.2}
	\centering
	\revise{
	\caption{Paired t-Test for our method with two baseline methods using Average Symmetric Distance.}
	\label{tab:test_2}
	\begin{tabular}{l|  c  c c c}
		\hline
        Methods&Site A & Site B &Site C &Overall\\
		\hline
        Joint &0.0273 &0.0018 &0.0330 &8.2e-5\\
        Separate &0.0192 &0.0187 &0.0362 &0.0001\\
		\hline
	\end{tabular}
	}
\end{table}
\vspace{-2mm}
\subsection{Analysis of Inter-site Heterogeneity}
To deal with multi-site data, it is essential to first quantitatively analyze the inter-site heterogeneity of datasets from different sites.
%
Following the experiment strategy to analyze domain shit in~\cite{dou2018unsupervised}, we first conducted cross-site validation among the three datasets. Specifically, we trained the \emph{Separate} models on each dataset and evaluated within and across different datasets. 
As shown in the top part of Table~\ref{tab:pre-processing}, the dice score of \emph{Separate} approach is relatively high when evaluating within the same dataset, while catastrophically drops when evaluating across different datasets. These cross-validation results clearly reveal the sensible discrepancy among different sites. 

Under multi-site learning scenario, we concern more about whether such discrepancy still matters when incorporating these data for network training, and whether careful pre-processing techniques could sufficiently alleviate the inter-site discrepancy.
%
\revise{We therefore analyse the effects of three different intensity normalization methods on the performance of \emph{Joint} approach, including 1) histogram matching to a target distribution (histogram); 2) mapping the intensities to range between [0,1] (scaled); and 3) mapping the intensities to have zero mean and unit variance (whitening). We also study the benefits of bias field correction and noise filtering techniques, based on the observation from~\cite{palumbo2011interplay,glocker2019machine} that these techniques play an important role in establishing an optimal pre-processing sequence to deal with multi-site imaging data. }

\revise{The results are listed in Table~\ref{tab:pre-processing}. We observe these three kinds of intensity normalization methods are all effective to improve the performance of \emph{Joint} approach, among which the whitening normalization performs the best and could improve the dice score by 3.15\% compared with \emph{Joint} model without intensity normalization. Fig.~\ref{fig:statistic} illustrates the intensity histograms before and after the whitening normalization, from which we could clearly see that whitening normalization is indeed helpful to harmonize the distribution of gray-scale values. We also notice from Table~\ref{tab:pre-processing} that the bias field correction and noise filtering could improve the performance over whitening normalization, while the improvement is limited and not consistent on different data sites. It is worth noting that even though conducting very careful pre-processing (Bias Field Correction - Noise Filtering - Whitening) as~\cite{palumbo2011interplay}, the \emph{Joint} approach cannot show explicit advantage over the \emph{Separate} approach with only whitening normalization, and still under-performs the \emph{Separate} model on Site B.} Similar observations on prostate segmentation have also been reported in other recent studies. 
Onofrey \etal~\cite{john2019generali} explore four different normalization methods, including scaled, Gaussian, quantile normalization, and histogram matching, 
to normalize prostate MRI images from two different sites.
They find that none of these methods achieves considerable improvements compared with \emph{Separate} models.

\revise{Experiments above show that carefully pre-processing is helpful but insufficient to fundamentally solve the data heterogeneity problem when incorporating multi-site data for network training.} This is probably because the inter-site heterogeneity not only comes from intensity variance, but is also due to other factors (\eg, coil use and resolution difference), which can not be addressed by the pre-processing techniques.
Meanwhile, these experiments also emphasize the interest and necessity of designing other effective approaches to learn robust representations from multi-site data to effectively improve the performance on each data site.
Note that in the following experiments, all the models are trained with whitening normalized data.
%
\vspace{-2mm}
\subsection{Effectiveness of Our Multi-site Learning Method}
\label{sec:effectiveourmethod}
In this section, we first compare our approach with the baseline \emph{Separate} and \emph{Joint} approaches, and then conduct comparison with the state-of-the-art approaches in multi-site learning. 
Before conducting the comparison, we evaluate the segmentation performance of our backbone network. 
To the best of our knowledge, Tian~\etal~\cite{tian2017super} and Rundo~\etal~\cite{rundo2019cross} achieved the state-of-the-art segmentation performance on the same three datasets that we utilize.
We directly reference their reported numbers in our paper to demonstrate that our implemented segmentation backbones are valid.
Specifically, in Table~\ref{tab:comparisons}, our \emph{Separate} model achieves higher performance than their reported performance. 

\subsubsection{Comparison with baseline settings}
\begin{figure*}[t]
	\centering
	\includegraphics[width=1.0\textwidth]{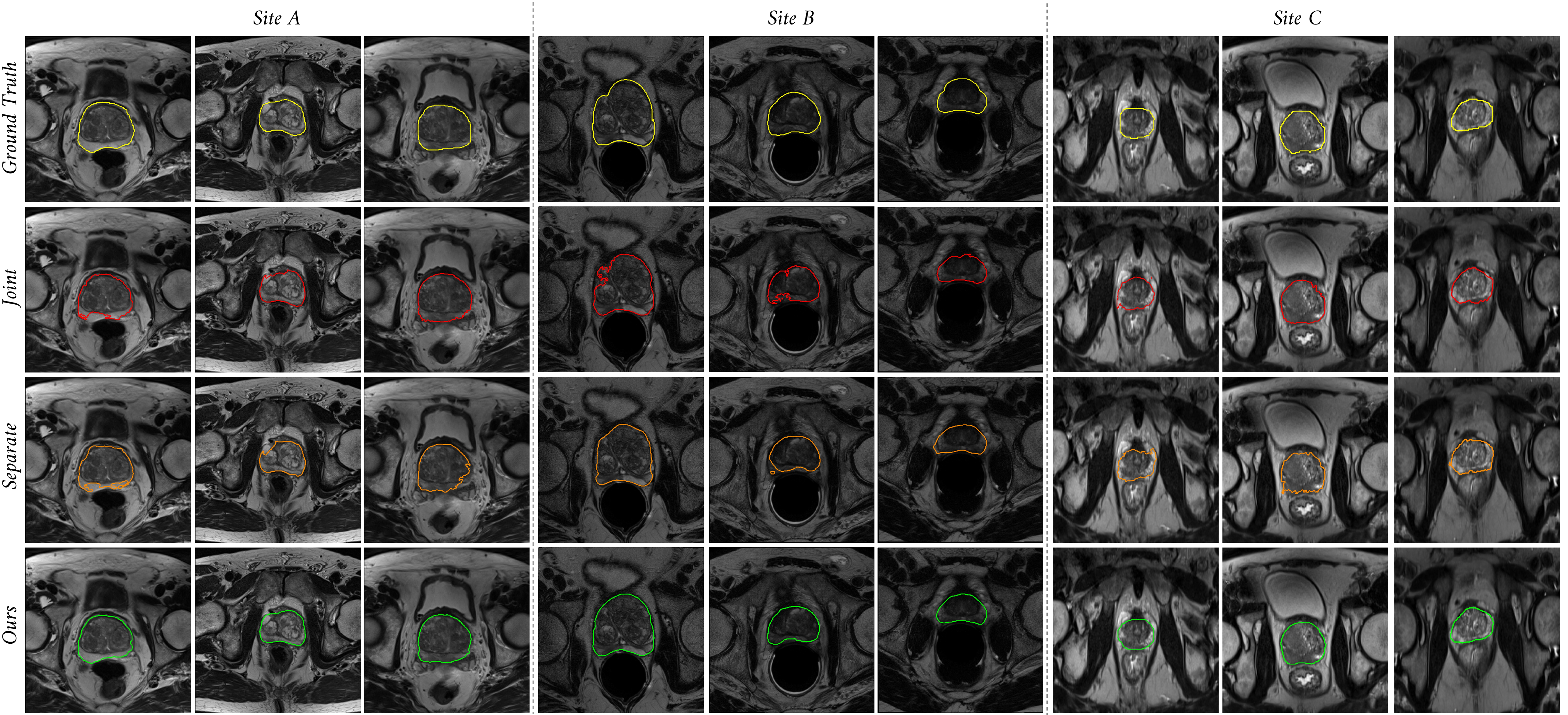}
	\vspace{-3mm}
	\caption{Qualitative segmentation results on the three heterogeneous datasets. From top to bottom are the ground truth, results of \emph{Joint} approach, \emph{Separate} approach and our approach, respectively. Image intensities are normalized before inputting to these networks.}
	\label{fig:visual_results}
	\vspace{-2mm}
\end{figure*}

%
In Table~\ref{tab:comparisons}, the \emph{Joint} approach under-performs \emph{Separate} approach in Site B and Site C, with both approaches trained using whitening normalized data. It is worthy to point out that when incorporating our DSBN layers, the performance of the joint network, \ie, \textit{DSBN}, outperforms the \emph{Separate} approach on all three sites consistently, \revise{in both Dice score and Average symmetric distance}. This result indicates that our designed DSBN layer is suitable and effective for tackling the inter-site discrepancy under multi-site learning scenario.
Moreover, by utilizing our learning approach to enhance the learning of shared kernels, the final method \textit{MS-Net} further gains considerable improvements on three sites compared with the \textit{DSBN} network, \revise{obtaining average segmentation Dice of 91.66\% and overall Average symmetric distance of 0.77mm}. 
These improvements highlight that our learning approach indeed helps to 
capture more robust representations with better performance on multiple heterogeneous datasets.
We show some qualitative segmentation examples in Fig.~\ref{fig:visual_results} for visual comparison. 
It is observed that: 1) compared with the \emph{Joint} model, the \emph{Separate} model produces relatively better results in site B, due to the large discrepancy between site B and the other two sites; 2) our approach produces more accurate segmentation mask and delineates the clear boundary for all three sites consistently.

\begin{figure}[t]
	\centering
	\revise{
	\includegraphics[width=0.48\textwidth]{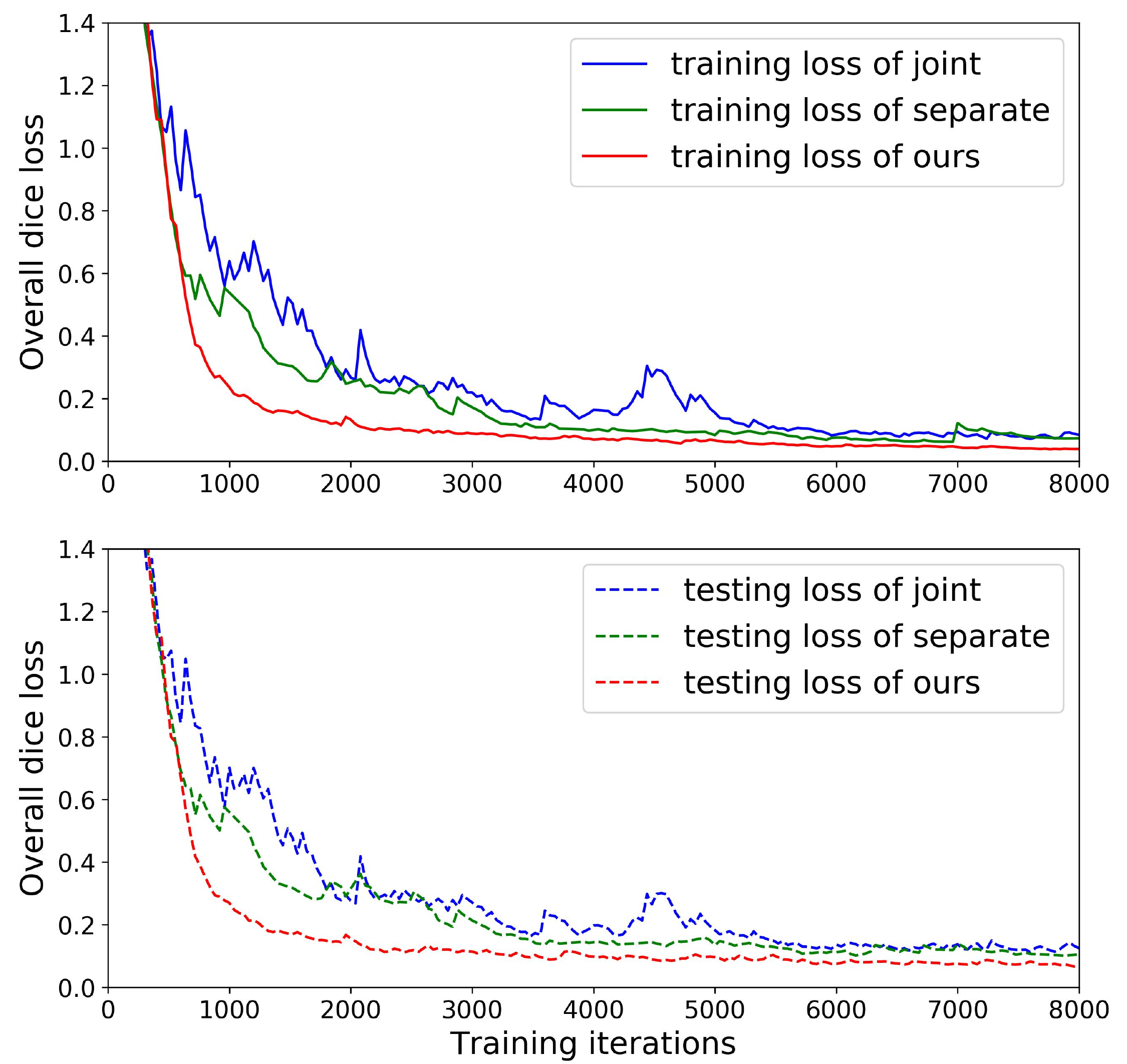}
	\vspace{-1mm}
	\caption{The training and testing dice loss summed over three data sites.}
	\label{fig:loss_curve}
	\vspace{-2mm}
	}
\end{figure}
To analyze whether the performance improvement of our method is significant, we conduct paired t-test for our method with the \emph{Joint} and \emph{Separate} models. 
We utilize \revise{dice score and average symmetric distance} as the evaluation measurement and set the significance level as 0.05. 
For each pair of comparison, we calculate the single-site p-value and overall p-value, respectively. The detailed results are shown in Table~\ref{tab:test} \revise{and Table~\ref{tab:test_2}}. 
All paired t-tests present p-value smaller than 0.05, demonstrating that our improvements compared with these baseline approaches are statistically significant.

\revise{We further compare the training and testing computational time of our method with the two baseline approaches. To compare the training computation cost, we visualize the overall dice loss curve of these approaches to observe their learning behavior, as illustrated in Fig.~\ref{fig:loss_curve}. Please note that in each training iteration, we feed three batches of images as the network input to fulfill the setting of multi-site learning, with one batch from each data site. The overall dice loss refers to the dice loss summed over the three batches of images. 
%
We observe that the \emph{Separate} approach converges faster than the \emph{Joint} approach, reflecting that the inter-site heterogeneity would affect the convergence speed when incorporating multi-site data for network training. Meanwhile, our approach converges faster and produces smoother loss curve than the \emph{Joint} and \emph{Separate} approaches, which indicates that the proposed method could effectively mitigate the data heterogeneity and improve the convergence speed by learning robust representation from multiple data sites. Furthermore, we also quantitatively evaluate the training time of these approaches. The \emph{Joint}, \emph{Separate} and \emph{Ours} approaches respectively take 5.5, 3.5 and 3 hours to converge, which is consistent with the observations above.}

\revise{We then evaluate the testing computational time, for which we randomly select three samples from each data site and compare the inference time of the three approaches on these samples. The 10 times average inference time for the \emph{Joint}, \emph{Separate} and \emph{Ours} approaches on these samples are 15.035s, 16.460s and 15.224s, respectively. Our approach takes similar inference time as the \emph{Joint} approach, which is reasonable as in testing phase, the DSBN layer only requires to choose the corresponding batch normalization layer for feature normalization, without adding additional computation cost. The \emph{Separate} models, even have the same architecture as the \emph{Joint} model, take more time for inference, which could be explained by that the \emph{Separate} approach requires to switch between different models when testing on different data sites. 
}

\subsubsection{Comparison with the state-of-the-art} 
To show the effectiveness of our method, we conduct comparison with the multi-site learning methods in medical image analysis area and the state-of-the-art multi-domain learning methods in natural image processing. 
Specifically, the compared methods include:
\begin{itemize}
    \item \textbf{USE-Net}~\cite{rundo2019use}: They incorporate Squeeze-and-Excitation blocks into Res-UNet to tackle the inter-site heterogeneity by channel-wise feature recalibration.
    
    \item \textbf{Dual-Stream}~\cite{valindria2018multimodal}: They propose a dual-stream architecture to learn the generic representations shared in CT and MRI. Following their setting, we share the bottleneck layers among three sites in our experiment.
    
    \item \textbf{Series-Adapter}~\cite{rebuff2017learning}: They propose the series adapter for multi-domain learning, which incorporates a designed adapter module in series into residual block to handle the variations among different visual domains.
    \item \textbf{Parallel-Adapter}~\cite{rebuff2018efficient}: They propose the parallel adapter, which incorporates the domain adapter in parallel with filter banks in residual block. The \emph{Parallel-Adapter} achieves the \textbf{state-of-the-art} classification accuracy for the joint learning task from 10 different visual domains, including ImageNet, CIFAR100, etc.
\end{itemize}

In this section, we take the \emph{Joint} approach as a reference to compare different multi-site learning methods.  
In Table~\ref{tab:comparisons}, \textit{USE-Net}~\cite{rundo2019use} achieves higher overall performance than the \emph{Joint} approach. However, the improvement is limited and it still under-performs the \emph{Separate} approach in site B, indicating that only increasing the network complexity and capacity is insufficient to tackle the inter-site heterogeneity and adequately learn the shared information from multi-site data. 
The \textit{Dual-Stream}~\cite{valindria2018multimodal} architecture designed for CT and MRI presents limited advantage over the \emph{Joint} approach in our experiment. 
Regarding the multi-domain learning methods,~\ie, \textit{Series-Adapter}~\cite{rebuff2017learning} and \textit{Parallel-Adapter}~\cite{rebuff2018efficient}, we observe the same phenomena as the original study that adding series or parallel convolutional kernels could effectively improve the performance, which demonstrates that domain-specific parameters are helpful to alleviate the inter-site heterogeneity.
%
However, different from the multi-domain learning approache which aims to establish a common network for handling different visual domains, here we further aim to explore the generic reprentation from multi-site data for improved prostate segmentation.
%
With the assistance of our learning approach, the proposed MS-Net outperforms the state-of-the-art method,~\ie,~\textit{Parallel-Adapter}, \revise{by 0.78\% in Dice score and 0.07mm in Average symmetric distance for the overall performance}. This result highlights the effectiveness of our approach in learning robust representation from multi-site data. 

\subsection{Ablation Analysis of Our Approach}
\subsubsection{MSKT with different loss ratio}
We study the effect of different hyper-parameter settings for $\alpha$ in Eq.~\eqref{eq:totalloss1}.
As shown in Table~\ref{tab:MSKT}, we gradually increase the ratio between knowledge transfer loss and the segmentation loss. 
\revise{It is observed that the MSKT could generally improve the segmentation performance when we range the loss ratio around 0.5, demonstrating the benefits of the regularization from transferred multi-site knowledge.}
Meanwhile, experiment shows that the ratio can not be set too high, only training the universal network with transferred knowledge ($\alpha$ = 1.0) can not guarantee an acceptable increase. This result indicates that the guidance from ground truth and auxiliary branches is complementary to each other, and both are indispensable to learn more robust representation from multi-site data.
In addition, we see that even without knowledge transfer from auxiliary branches ($\alpha = 0.0$), the universal network in our approach still achieves a higher performance than our approach without MSKT, \ie, \textit{DSBN}. This result shows that the multi-branch architecture can indeed perform as positive feature regularization to the universal network by jointly training the auxiliary branches and universal network. 

\begin{table}[t]
    \renewcommand\arraystretch{1.2}
	\centering
	\caption{Comparison of proposed approach with different knowledge transfer rate.}
	\label{tab:MSKT}
	\begin{tabular}{c| c| c  c  c | c}
		\hline
        Methods & $\alpha$ &Site A &Site B &Site C & Overall \\
        \hline
        MS-Net & 0.0 &91.23 &90.65 &91.54 &91.14\\
        MS-Net & 0.1 &91.56 &90.74 &91.72 &91.34\\
        MS-Net & 0.2 &91.30 &90.85 &91.82  &91.32\\
        MS-Net & 0.3 &91.20 &91.33 &91.91 &91.48\\
        MS-Net & 0.4 &\textbf{91.69} &91.20 &91.93  &91.61\\
        MS-Net & 0.5 &91.54 &91.24 &\textbf{92.18} &91.66\\
        MS-Net & 0.6 &91.53 &\textbf{91.43} &92.13 &\textbf{91.69}\\
        MS-Net & 0.7 &91.38 &91.35 &92.05 &91.59\\
        MS-Net & 0.8 &91.48 &90.67 &92.00 &91.38\\
        MS-Net & 0.9 &91.11 &90.94 &91.86 &91.30\\
        MS-Net & 1.0 &90.95 &90.71 &91.51 &91.06\\
        \hline
		Joint & - &90.69 &89.53 &90.55 &90.25\\
        DSBN  & - &90.98 &90.67 &91.07 &90.91\\
        Joint-MSKT &0.5 &90.99 &90.17 &91.29 &90.82  \\
        \hline
	\end{tabular}	
\end{table}

%
%

We also utilize MSKT to train a \emph{Joint} network and observe a 0.57\% (90.25\% vs. 90.82\%) dice improvement.
This improvement is smaller than the performance gain (0.75\%) by conducting MSKT on the DSBN network (see Section~\ref{sec:effectiveourmethod}). 
The reason may be that when utilizing DSBN to tackle the inter-site variations, the shared kernels gain more capacity to learn the generic representation from multi-site data. 
\subsubsection{Experiment with different backbone networks}
Our proposed approach is flexible to different network backbone designs and could be easily incorporated into other segmentation networks. 
We thus apply our approach on different backbone architectures, including Res-UNet~\cite{yu2017volumetric}, Dense-Unet~\cite{huang2017dense} and Mobile-Unet~\cite{howard2017mobile}.
The results of these experiments are presented in Table~\ref{tab:backbone}. 
Our approach increases the overall dice performance by 1.41\%, 1.27\% and 1.92\% for the three network backbones respectively,
which demonstrates the feasibility and general effectiveness of the proposed approach.
It is also observed that the Mobile-UNet produces relatively poor overall performance due to its lightweight network design compared with the other two networks. 
In that case, our approach brings a larger dice improvement, indicating that the regularization of transferred multi-site knowledge could play a more important role under the lightweight network backbone setting. 

\begin{table}[t]
    \renewcommand\arraystretch{1.2}
	\centering
	\caption{Comparison of proposed approach with different network backbones.}
	\label{tab:backbone}
	\begin{tabular}{l  c|c c  c | c}
		\hline
        Methods&Backbone &Site A &Site B &Site C &Overall\\
		\hline
        Joint &Res-UNet~\cite{yu2017volumetric}  &90.69 &89.53 &90.55 &90.25\\
        Ours &Res-UNet~\cite{yu2017volumetric} &91.54&91.24&92.18&91.66\\
        \hline
        Joint &Dense-UNet~\cite{huang2017dense} &90.77 &89.93 &90.41 &90.37\\
        Ours &Dense-UNet~\cite{huang2017dense} &91.59 &91.34 &92.00 &91.64\\
        \hline
        Joint &Mobile-UNet~\cite{howard2017mobile} &88.65 &88.45 &88.00 &88.36\\
        Ours &Mobile-UNet~\cite{howard2017mobile} &90.20 &90.49 &90.15 &90.28\\
		\hline
	\end{tabular}
\end{table}

%% file: discussion.tex
\section{Discussion}
\label{sec:discussion}
A variety of deep-learning approaches~\cite{tian2018psnet, wang2019deeply, cheng2017automatic} have achieved remarkable performance for automated prostate segmentation. 
These data-starving approaches commonly demand a large amount of training data for a high segmentation performance. Since it is much difficult to collect extensive training samples from a certain site (hospital) in real-world practice, it is meaningful to incorporate multi-site data for robust model training to alleviate the insufficiency of single-site samples. 
However, how to overcome the heterogeneous characteristic of multi-site data is a non-trivial problem.
Domain shift among multi-site data has long been a challenging problem in medical image analysis. 
Most previous works focus on unsupervised domain adaptation~\cite{dou2018unsupervised,chen2018semantic}, aiming to adapt the model trained on source site into a target site. 
These works typically maximize the performance on the target site and ignore the performance on the source site. 
Different from these previous works, in this paper, we study a different yet important problem, \ie, \textit{multi-site learning}, which aims at improving the performance on multiple sites simultaneously by learning more robust representations from multi-site data. 

We propose a novel framework to improve prostate segmentation on multi-site data by leveraging domain-specific batch normalization layers to compensate for the inter-site heterogeneity and a novel learning scheme to enhance the learning of shared kernels. 
Data used in our experiment are from three sites~\cite{bloch2019isbi, lemaitre2015i2cvb}, which presents visible heterogeneity and is similar to the real clinical circumstance. 
We do not use the popular dataset from PROMISE 12 Challenge~\cite{litjens2014promise}, since this dataset includes scans from four different sites without site information of each case, making it hard for us to evaluate the segmentation performance on each site separately. 
Extensive experiments have demonstrated that our approach consistently improves the performance on all sites by learning robust representation from multi-site data, outperforming the baseline approaches and state-of-the-art approaches for multi-site learning. 
%
\begin{table}[t]
    \renewcommand\arraystretch{1.2}
	\centering
    \revise{
	\caption{\small{Comparison on generalization ability of different approaches.}}
	\label{tab:generalization}
	\scalebox{1.00}{
	\begin{tabular}{l|  c c  c }
		\hline
         &Separate &Joint &Ours \\
		\hline
        Dice (\%) &57.53$\pm$11.36&73.14$\pm$17.94&70.68$\pm$12.92\\
		\hline
	\end{tabular}	
	}
	}
\end{table}

Although the good performance is achieved, the limitation of our method still exists.
In our method, the DSBN layer allocates the BN layer separately to collect more accurate statistic information for each data site and provides domain-specific trainable variables for better future normalization. While this design also brings some limitations during the inference phase. 
First, the site prior information of a testing sample need to be supplied during testing, so that our network could utilize the corresponding BN statistics for feature normalization. 
Fortunately, this prior could be obtained from the header information of a clinical data. 
\revise{In addition, the design of DSBN limits the generalization ability of our approach since when it comes an unseen data site, the corresponding statistic information and the trainable variables of that domain are unknown.} 

\reviseminor{In the context of multi-site learning, this work focuses on and is limited to improving the performance on internal data sites by leveraging multi-site data, and the performance on external data sites is not considered.} Even though the generalization is out of the scope, we try to provide some observations for the generalizability of our approach \emph{w.r.t.} the two baseline approaches. 
We construct an external testing set of 20 samples from Promise12 dataset~\cite{litjens2014promise}. For the evaluation of \emph{Ours} and \emph{Separate} approaches, we test the external set repeatedly using the BN layer (or separate model) of different data sites, and then average the results as the final predictions. The results are listed in Table~\ref{tab:generalization}.  We observe that the \emph{Joint} approach obtains the best performance on the external test set, showing that directly incorporating multi-site data for training helps to learn a widespread distribution. The \emph{Separate} approach presents poor generalization ability, which indicates that this approach is not suitable for generalization task. Meanwhile, our approach performs better than \emph{Separate} approach but is inferior to the \emph{Joint} approach, which is reasonable as we collect the statistic information for each data site separately to improve the performance on internal data sites, rather than learning a global distribution for the generalization purpose.

Regarding this limitation above, in the future, 
%
%
we will study how to effectively utilize multi-site data to improve the performance on both internal and external unseen data sites, which is more helpful to solve the clinical requirements, as well as a more challenging problem to be explored~\cite{dou2019domain}. Moreover, we also plan to investigate the multi-site learning problem in other medical imaging computing tasks,~\eg, diseases classification from chest radiograph. As reported in~\cite{kevin2019baseline}, aggregating multiple large chest radiograph datasets lead to performance degradation on all datasets more or less. Analyzing and solving the inter-site heterogeneity among million level chest radiographes is a promising extension as our future work.

%% file: conclusion.tex
\section{Conclusion}
\label{sec:conclusion}
We propose a novel multi-site network (i.e., MS-Net) for improved prostate segmentation by learning the shared knowledge from multiple heterogeneous datasets. Our framework explicitly tackles the inter-site heterogeneity by utilizing DSBN layer and can also capture more robust representations from multi-site data with the assistance of transferred multi-site knowledge. 
Our method gracefully addresses the inter-site heterogeneity of clinical prostate MRIs for robust model training. 
Extensive experiments on three heterogeneous datasets demonstrate the superiority of our approach compared with the baseline approaches and other state-of-the-art methods. In addition, our proposed approach is a general strategy that could be applied to other tasks under multi-site learning scenarios in real-world clinical practice.